\documentclass[conference,10pt,letterpaper]{IEEEtran} 

\usepackage[utf8]{inputenc} 
\usepackage[T1]{fontenc}
\usepackage[cmex10]{amsmath}
\usepackage{amssymb,amsfonts,amsthm}
\usepackage{algorithmic}
\usepackage{algorithm}
\usepackage{array}
\usepackage[caption=false,font=normalsize,labelfont=sf,textfont=sf]{subfig}
\usepackage{textcomp}
\usepackage{stfloats}
\usepackage{url}
\usepackage{verbatim}
\usepackage{graphicx}
\usepackage{cite}
\usepackage{xcolor}
\usepackage{braket}
\usepackage[hidelinks]{hyperref}
\usepackage{cleveref}
\usepackage{gensymb}
\usepackage{orcidlink}

\usepackage{etoolbox}
\newtoggle{isISIT}

\crefformat{equation}{(#2)#1(#3)}
\crefrangeformat{equation}{(#3#1#4)--(#5#2#6)}

\crefname{equation}{}{} 
\creflabelformat{equation}{(#2\textup{#1}#3)} 

\def\BibTeX{{\rm B\kern-.05em{\sc i\kern-.025em b}\kern-.08em
    T\kern-.1667em\lower.7ex\hbox{E}\kern-.125emX}}

\newcommand{\ketbra}[2]{\ket{#1}\bra{#2}}
\newcommand{\pfail}{p_\text{f}}
\newcommand{\bobiqubit}{\hat{\tau}_{\psi}^{B_i}}

\DeclareMathOperator{\supp}{supp}

\newtheorem{theorem}{Theorem}

\DeclareMathOperator{\tr}{tr}

\allowdisplaybreaks

\begin{document}

\title{Achievability of Covert Quantum Communication}
\author{
\IEEEauthorblockN{Evan J.~D.~Anderson\IEEEauthorrefmark{1}, Michael S.~Bullock\IEEEauthorrefmark{2}, Filip Rozp\k{e}dek\IEEEauthorrefmark{3}, and Boulat A.~Bash\IEEEauthorrefmark{2}\IEEEauthorrefmark{1}}
\IEEEauthorblockA{\IEEEauthorrefmark{1}Wyant College of Optical Sciences, University of Arizona
Tucson, AZ, USA \\
 Email: \{ejdanderson, boulat\}@arizona.edu}
\IEEEauthorblockA{\IEEEauthorrefmark{2}Electrical and Computer Engineering Department, University of Arizona, Tucson, AZ, USA \\ 
Email: bullockm@arizona.edu}
\IEEEauthorblockA{\IEEEauthorrefmark{3}College of Information \& Computer Sciences, University of Massachusetts, Amherst, MA, USA\\
Email: frozpedek@umass.edu}
\thanks{Full paper, including appendix, is available at \cite{andersonAchievabilityCovertQuantum2025}}
\thanks{This material is based upon work supported by the National Science Foundation under Grants No. CCF-2006679 and EEC-1941583.}
}

\maketitle

\begin{abstract}
\iftoggle{isISIT}{THIS PAPER IS ELIGIBLE FOR THE STUDENT PAPER AWARD. }{}
We explore covert communication of qubits over an arbitrary quantum channel. Covert communication conceals the transmissions in the channel noise, ensuring that an adversary is unable to detect their presence. We show the achievability of a \emph{square root law} (SRL) for quantum covert communication similar to that for classical: $M(n)\propto\sqrt{n}$ qubits can be transmitted covertly and reliably over $n$ uses of a general quantum channel.  We lower bound $M(n)$ with and without assistance from a two-way covert classical channel. In the former case, we quantify the number of classical covert bits sufficient for our protocol.
\end{abstract}

\section{Introduction}

Covert, or low probability of detection/intercept (LPD/LPI), communication conceals the transmissions in the channel noise, ensuring that an adversary is unable to detect their presence. Over the last decade, the fundamental limits of covert communication were explored for classical  \cite{bash15covertcommmag,bash13squarerootjsac,bloch15covert, wang15covert} and classical-quantum channels \cite{bash15covertbosoniccomm, bullock20discretemod, gagatsos20codingcovcomm, azadeh16quantumcovert-isitarxiv, bullock2025fundamentallimitscovertcommunication}. Covert communication over these channels is governed by the \textit{square root law} (SRL): $L_{\rm c}\sqrt{n}$ covert bits are reliably transmissible over $n$ channel uses for a channel-dependent constant $L_{\rm c}>0$. 

Recently, we extended these results to quantum covert communication over the lossy thermal-noise bosonic channel \cite{anderson2024covert-qce}.
We found an SRL similar to classical: $L_{\rm q}\sqrt{n}$ covert qubits are reliably transmissible over $n$ channel uses for a channel-dependent constant $L_{\rm q}>0$. 
The bosonic channel models quantum-mechanical communication in optical fiber, free space optics (FSO), microwave, and radio-frequency (RF) domains. However, notwithstanding its practical utility, limiting the analysis to a specific channel substantially restricts its application.
Thus, in this work we investigate achievability of the SRL for covert quantum communication over an arbitrary quantum channel.

Achievability of covert quantum communication has been previously studied in the context of quantum key distribution (QKD) \cite{arrazolaCovertQuantumCommunication2016, tahmasbi19covertqkd,tahmasbi20bosoniccovertqkd-jsait,tahmasbi20covertqkd}. Here, we are interested in quantifying the number of covert qubits that can be reliably transmitted between parties that already possess covert classical resources such as a pre-shared classical secret and a covert classical channel. Hence, we take a direct approach and derive achievability proofs for quantum covert communication over a general quantum channel under some reasonable restrictions on the adversary's output state.

Similar to our prior result on covert quantum communication over the bosonic channels \cite{anderson2024covert-qce}, we adapt the analysis from covert classical-quantum channels \cite{bash15covertbosoniccomm, bullock20discretemod, gagatsos20codingcovcomm, azadeh16quantumcovert-isitarxiv, bullock2025fundamentallimitscovertcommunication}. 
However, the results here apply to all finite-dimensional quantum channels that allow covert communication per criteria in \cite[Th.~5]{bullock2025fundamentallimitscovertcommunication} and many infinite-dimensional ones, including the bosonic channel explored in \cite{anderson2024covert-qce}.
In fact, our new approach improves the achievable constant in the SRL for bosonic quantum covert communication over that in \cite{anderson2024covert-qce}.

The rest of this paper is organized as follows: in Section~\ref{sec:preliminaries} we state the mathematical preliminaries as well as the system and channel models. In Section~\ref{sec:primary-results} we provide our results: the achievable lower bounds for covert quantum communication over a general quantum channel with and without a classical covert channel. Finally, in Section~\ref{sec:discussion} we discuss the implications of our results and describe areas of future research.

\iftoggle{isISIT}{The full paper, including the appendix, is available at \cite{andersonAchievabilityCovertQuantum2025}.}{}

\section{Preliminaries}
\label{sec:preliminaries}

\subsection{System Model}
\label{subsec:systemmodel}

\begin{figure*}[htb]
\centering
\includegraphics[width=\textwidth]{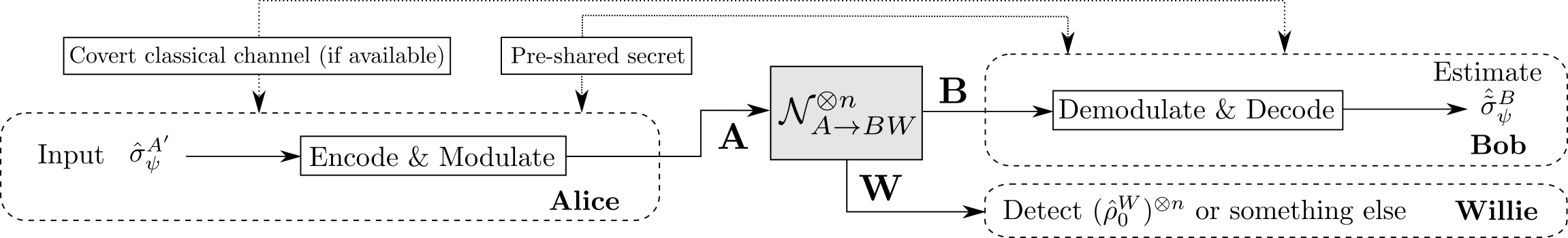}
\caption{System model for covert quantum communication. Alice has an input $\hat{\sigma}^{A^\prime}_\psi$  or an innocent state $\hat{\rho}_{\textrm{inn}}^A=\ketbra{\textrm{inn}}{\textrm{inn}}^A$ and uses an encoder and modulation before transmission through the $n$ uses of channel $\mathcal{N}_{A\to BW}$. Bob receives a state and demodulates then decodes to generate an estimate $\hat{\tilde{\sigma}}^B_\psi$ of the original message. Willie attempts to detect whether the innocent state was input or something else. When the innocent state is input on $n$ channel uses, he detects $(\hat{\rho}_0^W)^{\otimes n}$. Alice and Bob utilize a pre-shared secret, and, potentially, a covert classical communication channel to aid in covert quantum communication.}
\label{fig:system}
\end{figure*}

Consider covert communication setting described in Fig.~\ref{fig:system}.
Alice wishes to transmit a quantum state $\ket{\psi}^{A^\prime}$ with density matrix $\hat{\sigma}_{\psi}^{A^\prime}=\ketbra{\psi}{\psi}^{A^\prime}$ to Bob without being detected by an adversarial warden Willie. This state is encoded by a map $\mathcal{E}_{A^\prime\to A^n}$ that outputs state $\hat{\rho}_{\psi}^{A^n}=\mathcal{E}_{A^\prime\to A^n}\left(\hat{\sigma}_{\psi}^{A^\prime}\right)$ occupying $n$ qubits.  In turn, $\hat{\rho}_{\psi}^{A^n}$ is modulated and transmitted over $n$ uses of a quantum channel  $\mathcal{N}_{A\to BW}$.  The modulator and the channel act independently on the $n$ states of each system $A$. Thus, Bob and Willie receive $\hat{\rho}^{B^n}_\psi \equiv\tr_{W}\left(\mathcal{N}^{\otimes n}_{A\to BW}(\hat{\rho}^{A^n}_{\psi})\right)$ and $\hat{\rho}^{W^n}_\psi \equiv\tr_{B}\left(\mathcal{N}^{\otimes n}_{A\to BW}(\hat{\rho}^{A^n}_{\psi})\right)$, respectively. 
We assume that Alice and Bob pre-share a classical secret, as is standard in covert communications \cite{bash15covertbosoniccomm, bullock20discretemod, gagatsos20codingcovcomm, azadeh16quantumcovert-isitarxiv, bullockCovertCommunicationClassicalQuantum2023, bash15covertcommmag,bash13squarerootjsac,bloch15covert, wang15covert, anderson2024covert-qce}.
We also investigate the impact of the availability of a classical covert channel.

\subsection{Hypothesis Testing and Covertness}\label{subsec:hypothesis-testing}
We assume that Willie has complete knowledge of the system in Fig.~\ref{fig:system}, except for Alice and Bob's pre-shared secret.
Willie must determine from his channel output whether Alice is using the channel (hypothesis $H_1$) or not (hypothesis $H_0$).
Let $\hat{\rho}^W_0\equiv\tr_{B}\left(\mathcal{N}_{A\to BW}(\hat{\rho}_{\textrm{inn}}^{A})\right)$ where $\hat{\rho}_{\textrm{inn}}^A$ denotes the ``innocent'' input, indicating that Alice is not transmitting (e.g., vacuum input into a bosonic channel as in \cite{anderson2024covert-qce}). 
Thus, over $n$ channel uses, Willie's output state under $H_0$ is $\left(\hat{\rho}^W_0\right)^{\otimes n}=\tr_{B}\left(\mathcal{N}^{\otimes n}_{A\to BW}\left((\hat{\rho}_{\textrm{inn}}^{A})^{\otimes n}\right)\right)$.
Under $H_1$, Willie receives $\hat{\rho}^{W^n}=\sum_\psi p(\psi)\hat{\rho}^{W^n}_{\psi}$, where $p(\psi)$ is an arbitrary distribution representing Willie's uncertainty about Alice's message, and $\hat{\rho}^{W^n}_{\psi}$ is the non-innocent output state at Willie during a transmission from Alice over $n$ channel uses defined in Section \ref{subsec:systemmodel}. 
Willie uses arbitrary quantum resources to discriminate between $H_0$ and $H_1$, including fault-tolerant quantum computers, perfect quantum measurement, and ideal quantum memories.

Assuming equal priors, i.e., $P(H_1)=P(H_0)=\frac{1}{2}$, Willie's  probability of error is 
    $P_e^W=\frac{P_{FA}+P_{MD}}{2}$,
with probability of false alarm $P_{FA}=P(\text{choose } H_1|H_0 \text{ true})$ and probability of missed detection $P_{MD}=P(\text{choose } H_0|H_1 \text{ true})$. $P_e^W\leq \frac{1}{2}$ is the trivial upper bound Willie can achieve by using a random decision device. Thus, Alice and Bob try to ensure that Willie's minimum probability of error is close to that of this ineffective device. Formally, they seek
    $P_e^W\geq\frac{1}{2}-\delta,$
where $\delta>0$ quantifies the desired level of covertness. Willie's minimum probability of error is bounded by trace distance between his output states under each hypothesis as  \cite[Sec.~9.1.4]{wilde16quantumit2ed}:
\begin{align}
    P_e^W\geq\frac{1}{2}-\frac{1}{4}\left\|\hat{\rho}^{W^n}-\left(\hat{\rho}_0^{W}\right)^{\otimes n}\right\|_1,
\end{align}
where $\|\cdot\|_1$ is the trace norm. 
The trace distance is often mathematically unwieldy. Conveniently, the quantum relative entropy (QRE) 
 $D\left(\hat{\rho}\middle\|\hat{\sigma}\right) \equiv \tr\left(\hat{\rho}\log\hat{\rho} - \hat{\rho}\log\hat{\sigma}\right)$ is additive over product states, and upper bounds the trace distance via the quantum Pinsker's inequality \cite[Th. 11.9.1]{wilde16quantumit2ed}:
\begin{align}
\frac{1}{4}\left\|\hat{\rho}^{W^{n}}-\left(\hat{\rho}_0^W\right)^{\otimes n}\right\|_1 \leq \sqrt{\frac{1}{8} D\left(\hat{\rho}^{W^{n}}\middle\|\left(\hat{\rho}_0^W\right)^{\otimes n}\right)}.\label{eq:pinskers}
\end{align}
We employ QRE as our covertness criterion, as is standard in both classical \cite{bloch15covert, wang15covert} and quantum \cite{bullock20discretemod, gagatsos20codingcovcomm, azadeh16quantumcovert-isitarxiv, bullock2025fundamentallimitscovertcommunication, anderson2024covert-qce} analyses.
Formally, we call a communication scheme $\delta_{\rm QRE}$-covert if 
\begin{align}
D\left(\hat{\rho}^{W^{n}}\middle\|\left(\hat{\rho}_0^W\right)^{\otimes n}\right) \leq \delta_{\rm QRE} \label{eq:covert-req}
\end{align}
for $\delta_{\rm QRE}>0$.

\noindent{\bf Assumptions on Willie's output states:} For a reliable covert communication scheme to be possible, 
\begin{align}
    \supp \hat{\rho}^{W^n}_{\phi} \subseteq \supp \left(\hat{\rho}^W_0\right)^{\otimes n} \label{eq:supports-covert-possible}
\end{align} for each non-innocent input state $\ket{\phi}\in\mathcal{H}^{A^\prime}$\cite[Th.~5]{bullock2025fundamentallimitscovertcommunication}. 

Further, in the case that Willie's output systems are infinite-dimensional, let $\hat{H}^{W}$ be a Hamiltonian for the output at Willie of a single channel use. We assume $\hat{H}^{W}$ satisfies the Gibbs hypothesis \cite[Sec.~IV]{winter16cont}, and that the innocent state $\hat{\rho}^W_0$ with finite energy $E_0$ is of Gibbs form \cite[Sec.~IV]{winter16cont}: 
\begin{align}
    \hat{\rho}^W_0 = \frac{1}{Z(\beta(E_0))}e^{-\beta(E_0) \hat{H}^W}\label{eq:rho0-Gibbs-form},
\end{align}
where $Z(\beta)=\tr\left(e^{-\beta\hat{H}^W}\right)$, $\beta(E_0)$ is the solution to the equation $\tr\left(e^{-\beta \hat{H}^W}(\hat{H}^W-E_0)\right)=0$. Finally, we assume that for the output state $\hat{\rho}^{W}_{\hat{\pi}} = \mathcal{N}_{A\to W}(\hat{\pi})$ corresponding to maximally mixed input state $\hat{\pi}=\frac{\hat{I}}{2}$ and for system Hamiltonian spectral decomposition $\hat{H}^{W}=\sum_{k} \lambda_{k} \hat{\Pi}_{k}$, we have  
\begin{align}
\tr\left(\left(\hat{\rho}^{W}_{\hat{\pi}}\right)^2 \hat{\Pi}_{k}\right) = o\left(\frac{e^{-\beta(E_0)\lambda_{k}}}{k}\right)  ,\label{eq:inf-chi-square-assumption}
\end{align}
where $f(k)=o(g(k))$ indicates $g(k)$ is an asymptotic upper bound on $f(k)$ such that  $\lim_{k\to\infty} \frac{f(k)}{g(k)} = 0$.

Unlike \eqref{eq:supports-covert-possible}, our assumptions in \cref{eq:rho0-Gibbs-form,eq:inf-chi-square-assumption} are unnecessary for covert communication to be possible in general (consider a trivial channel such that $\hat{\rho}^{W^n}_\phi$ = $\left(\hat{\rho}^W_0\right)^{\otimes n}$ for all input states $\ket{\phi}\in\mathcal{H}^{A^\prime}$) and arise mainly due to technical convenience in the covertness analysis component of the achievability proof. However, \eqref{eq:rho0-Gibbs-form} is the state that maximizes the von Neumann entropy for a given finite energy $E_0$, and corresponds to the output state in the practical scenario where the Alice to Willie channel is lossy thermal-noise bosonic channel and her innocent input state is vacuum \cite{anderson2024covert-qce}.  Furthermore, the maximally-mixed state in both single- and dual-rail qubit encodings for the bosonic channel in \cite{anderson2024covert-qce} satisfy \eqref{eq:inf-chi-square-assumption}.


\begin{figure*}[htb]
\centering
\includegraphics[width=\textwidth]{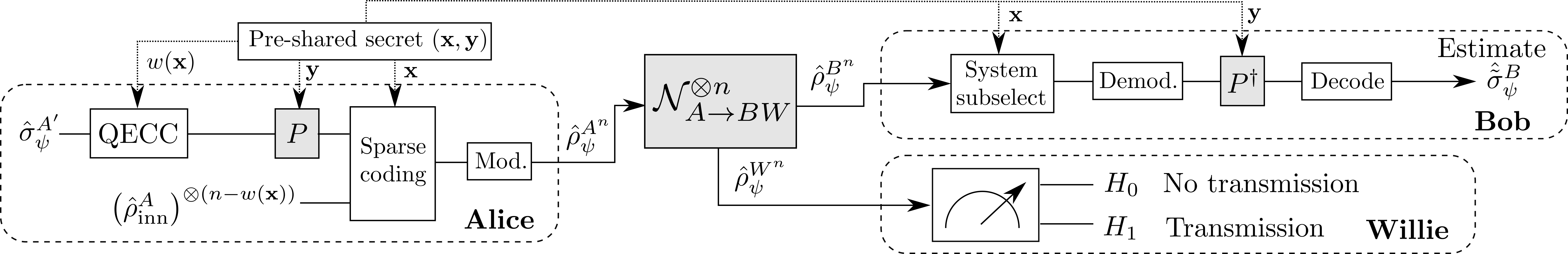}
\caption{Construction of achievable covert quantum communication over quantum channel $\mathcal{N}_{A\to BW}$. Given a pre-shared secret $\mathbf{x}$, Alice applies a QECC corresponding to the number $w(\mathbf{x})$ of non-innocent output states.
This is followed by the application  on the output qubit states of a pre-shared secret
 sequence $\mathbf{y}$ of $w(\mathbf{x})$ Pauli gates. 
She then applies sparse coding that ``spreads'' these $w(\mathbf{x})$ non-innocent states across $n$ channel uses by inserting innocent input states according to locations in $\mathbf{x}$.
Alice then transmits the modulated resulting state $\hat{\rho}^{A^n}_\psi$. Bob sub-selects the systems containing the non-innocent states using $\mathbf{x}$ and demodulates by projecting them into the qubit basis. He then inverts the Pauli twirling operation using $\mathbf{y}$ and decodes to obtain an estimate $\hat{\tilde{\sigma}}^{B^\prime}_\psi$ of the message. Willie receives $\hat{\rho}^{W^n}_\psi$ and performs an optimal hypothesis test on whether transmission occurred.}
\label{fig:construction-no-locc}
\end{figure*}

\subsection{Reliable Quantum Communication}
A primary challenge in quantum communication is the handling of errors introduced by noisy channels. Given a quantum state $\hat{\rho}^{B^n}_\psi$ received from Alice through a noisy channel, Bob obtains an estimate $\hat{\tilde{\sigma}}^{B^\prime}_{\psi}$ of $\hat{\sigma}^{A^\prime}_{\psi}$. We call our system \emph{reliable} if, for any $\epsilon >0$, there exists a sufficiently large block size (number of channel uses) $n$ and an encoding/decoding strategy such that $\|\hat{\sigma}^{A^\prime}_{\psi}-\hat{\tilde{\sigma}}^{B^\prime}_{\psi}\|_1 < \epsilon$.

One may also use the fidelity metric of the two states via its relation to trace distance using the Fuchs-van de Graaf inequalities \cite{fuchsCryptographicDistinguishabilityMeasures1999}. Additionally, vanishing expected error probability may be employed as in achievability proofs of the hashing bound \cite[Sec. 24.6.3]{wilde16quantumit2ed} for Pauli channels and the hashing method \cite{bennet96qec} for entanglement distillation.

\section{Results}\label{sec:primary-results}

Here, we provide achievable lower bounds on reliable covert quantum communication with and without an available two-way covert classical channel, beginning with the latter.
We denote by $[a]^+=\max(a,0)$, and the  Shannon entropy associated with probability vector $\vec{p}$ by $H(\vec{p})=-\sum_{p_i\in\vec{p}} p_i\log(p_i)$.

\subsection{Covert Quantum Channel Without Classical Assistance}
\begin{theorem}
\label{thm:lower-bound}
For $n$ large enough and arbitrary $\vartheta>0$, $M(n)\geq(1-\vartheta)\sqrt{n}c_{\mathrm{q}}R\sqrt{\delta_{\text{QRE}}}$  qubits can be transmitted reliably and covertly over $n$ uses of an arbitrary quantum channel that acts independently on each channel use, where $c_\mathrm{q}$ is defined in \eqref{eq:c_cov}, $\delta_{\text{QRE}}$ is the QRE-covertness constraint, and $R\geq\left[1-H(\vec{p})\right]^+$ is the constant achievable rate of reliable qubit transmission per channel use, with $\vec{p}$ defined in \cref{eq:pvec,eq:pvec-i,eq:pvec-xyz}. 
\end{theorem}
\begin{IEEEproof}
 \noindent{\bf Construction and reliability:} The construction of the achievable model is in Fig.~\ref{fig:construction-no-locc}.
To ensure \eqref{eq:covert-req}, Alice employs the sparse signaling approach from  \cite{tahmasbi21signalingcovert}:
 let $\mathbf{x}\equiv\{x_i, i=1,\ldots, n\}$ be a binary sequence indicating the selected channel uses for non-innocent transmission, with $x_i=1$ corresponding to qubit input from Alice on the $i^{\text{th}}$ channel use, and $x_i=0$ to innocent input. Thus, $w(\mathbf{x})\equiv\sum_{i=1}^{n}x_i$ is the number of non-innocent inputs for a given $\mathbf{x}$.
 For an arbitrary $\vartheta>0$, define $\mathcal{A}\equiv\left\{\mathbf{x}: \left|q-\frac{1}{n}w(\mathbf{x})\right|\leq\vartheta\right\}$
as the set containing length-$n$ binary sequences whose normalized weight is close to $q$.
Let 
\begin{align}
    p_X(x)&=\{q \text{~if~} x=1; 1-q \text{~if~} x=0\} \label{eq:p_X}.   
\end{align}
Denote $p(\mathcal{A})=\sum_{\mathbf{x}\in\mathcal{A}}\prod_{i=1}^{n}p_X(x_i)$, and:
\begin{align}
    p_{\mathbf{X}}\left(\mathbf{x}\right) & \equiv \left\{\frac{\prod_{i=1}^{n}p_X(x_i)}{p(\mathcal{A})} \text{~if~} \mathbf{x}\in\mathcal{A}; 
         0 \text{~if~} \mathbf{x}\notin\mathcal{A}\right\}.\label{eq:p_vecX}
\end{align}
Alice and Bob choose the channel uses for transmitting qubits by randomly sampling $\mathbf{x}\in\mathcal{A}$ using $p_{\mathbf{X}}$.
Their choice $\mathbf{x}$ comprises part of the classical pre-shared secret in Fig.~\ref{fig:construction-no-locc}. 

Alice and Bob further generate a set $\{c_i\}_{i\in w(\mathcal{A})}$ of $|w(\mathcal{A})|= \lceil 2 \vartheta q n \rceil$ non-covert quantum error correction codes (QECCs), where   $w(\mathcal{A})\equiv\{w(\mathbf{x}) : \mathbf{x}\in\mathcal{A}\}$.
For a given number of non-innocent states available $w(\mathbf{x})$, the non-covert QECC $c_{w(\mathbf{x})}$ maps the input state $\ket{\psi}
\rightarrow|c_{w(\mathbf{x})}(\psi)\rangle^{A^{w(\mathbf{x})}}$. We extend these non-covert QECCs to a covert QECC $c$, which maps $(\mathbf{x},|\psi\rangle) \rightarrow|c(\psi, \mathbf{x})\rangle^{A^n}$, with $|c_{w(\mathbf{x})}(\psi)\rangle^{A^{w(\mathbf{x})}}=\operatorname{tr}_{\mathbf{x}^c}\left(|c(\psi, \mathbf{x}\rangle)^{A^n}\right)$ corresponding to the non-covert QECC mapping for a given $w(\mathbf{x})$, and $\operatorname{tr}_{\mathbf{x}^c}$ indicates tracing over the innocent inputs defined by choice of $\mathbf{x}$. Given $\mathbf{x}$, the covert QECC can be thought of as the sparse encoding of the corresponding non-covert QECC of block length $w(\mathbf{x})$, with innocent states injected in each of the systems $\{A_i : x_i=0\}$.
Hence, the systems occupied by innocent states do not contribute to the error-correcting capabilities of the code but are utilized for covertness.

Alice and Bob also secretly choose a sequence $\mathbf{y}$ indicating $w(\mathbf{x})$ Pauli gates sampled uniformly at random from $\mathcal{P}\equiv\{\hat{I},\hat{X},\hat{Y},\hat{Z}\}$. Alice applies these to the states occupying the selected $w(\bf{x})$ input systems. This is the first part of what is known as Pauli twirling \cite{emersonSymmetrizedCharacterizationNoisy2007}. Alice then modulates and transmits $\hat{\rho}^{A^n}_\psi$ through the channel $\mathcal{N}_{A\to BW}$. Bob receives state $\hat{\rho}^{B^n}_\psi$ and uses $\mathbf{x}$ to subselect the state $\hat{\rho}^{B^{w(x)}}_\psi = \tr_{\mathbf{x}^c}(\hat{\rho}^{B^n}_\psi)$, which represents the state occupied by output systems of the non-covert QECC.

Denote $\hat{\rho}_{\psi}^{B_i}$ as the state of the $i^\text{th}$ subsystem in $\hat{\rho}^{B^{w(x)}}_\psi$. The state of each subsystem is demodulated by projecting it onto the qubit basis through the application of $\hat{\Pi} = \ketbra{0}{0}+\ketbra{1}{1}$. Projection is a probabilistic process with the probability of failure $\pfail$ that depends on the structure of $\mathcal{N}_{A\to B}$. When Bob fails to project, he replaces the state with the maximally mixed state $\hat{\pi}=\frac{\hat{I}}{2}$. This process yields a state in the qubit basis of $(1-\pfail)\hat{\Pi}\hat{\rho}_\psi^{B_i}\hat{\Pi}^\dagger + \pfail \hat{\pi} = (1-\pfail)\bobiqubit+\pfail\hat{\pi}$ where $\bobiqubit$ is the projected state in the $i^\text{th}$ system. This represents a depolarizing channel for $\bobiqubit$. Bob then uses pre-shared $\mathbf{y}$ to apply the appropriate sequence of Pauli gates, completing the Pauli twirling process. Pauli twirling by Alice and Bob on their qubit states guarantees a Pauli noise channel with channel parameters $\vec{q}_\text{tw}=\left(q_I,q_X,q_Y,q_Z\right)$ on $\bobiqubit$ \cite{emersonSymmetrizedCharacterizationNoisy2007}. The parameters in $\vec{q}_{\text{tw}}$ may be derived in the standard way through the Choi matrix resulting from the application of $\mathcal{N}_{A\to B}$ followed by projection $\hat{\Pi}$ (e.g., see \cite{rozpedekOptimizingPracticalEntanglement2018}). 

Recall a Pauli channel, $\mathcal{E}^{\vec{p}}_{\mathcal{P}}$, maps input state $\hat{\rho}$ as follows:
$\mathcal{E}^{\vec{p}}_{\mathcal{P}}(\hat{\rho})=p_I \hat{I} \hat{\rho} \hat{I} + p_X \hat{X} \hat{\rho} \hat{X} + p_Y \hat{Y} \hat{\rho} \hat{Y} + p_Z \hat{Z} \hat{\rho} \hat{Z}$ for $\vec{p}=(p_I,p_X,p_Y,p_Z)$. A depolarizing channel is given by $\mathcal{E}^{\lambda}_{\rm dep}(\hat{\rho})=\mathcal{E}^{\vec{p}_\text{dep}(\lambda)}_{\mathcal{P}}(\hat{\rho})$ with depolarizing parameter $\lambda$ and $\vec{p}_\text{dep}(\lambda) = \left(1-\frac{3\lambda}{4},\frac{\lambda}{4},\frac{\lambda}{4},\frac{\lambda}{4}\right)$. Hence the combination of the depolarizing channel from projection to the qubit basis and Pauli channel generated by twirling yields a combined Pauli channel,
\begin{align}
    \mathcal{E}_{\text{dep}}^{\pfail}\left(\mathcal{E}^{\vec{q}_\text{tw}}_{\mathcal{P}}\left( \bobiqubit\right ) \right) = \mathcal{E}_{\mathcal{P}}^{\vec{p}}\left(\bobiqubit\right)
\end{align}
where 
\begin{align}
    \vec{p}&=(p_I, p_X, p_Y, p_Z) \;\; \text{with} \label{eq:pvec}\\
    p_I &= (1-\frac{3}{4}\pfail)q_I \;\; \text{and} \label{eq:pvec-i} \\ 
    p_j &=(1-\frac{3}{4}\pfail)q_j+\frac{1}{4}\pfail  \;\; \text{for} \;\; j=X,Y,Z \label{eq:pvec-xyz}.
\end{align}
For every $w(\mathbf{x})\in w(\mathcal{A})$, the hashing bound guarantees the existence of a QECC with an achievable rate of $R=1-H(\vec{p})$ \cite[Sec. 24.6.3]{wilde16quantumit2ed}.

\noindent{\bf Covertness analysis:}
The construction procedure, channel parameters, value of $q$, the covert QECC, and the time of transmission are known to Willie.

The sequence of random Pauli operations Alice applies to her encoded state is unknown to Willie. Therefore, from Willie's perspective, Alice's average non-innocent input state given $\mathbf{x}$ and any $\ket{\psi}$ is:
\begin{align}
    \hat{\rho}^{A^{w(\mathbf{x})}} =\mathcal{E}^{1^{\otimes w(\mathbf{x})}}_{\rm dep}\left(\ketbra{c({\psi})}{c({\psi)}}^{A^{w(\mathbf{x})}}\right) = \left(\hat{\pi}\right)^{\otimes w(\mathbf{x})}.
\end{align}
where $\mathcal{E}_{\rm dep}^{1^{\otimes w(\mathbf{x})}}(\cdot)$ is the completely depolarizing channel with $\lambda=1$ over $w(\mathbf{x})$ non-innocent input states.
Thus, for every $\ket{\psi}$, Willie observes the mixture of product states given by
\begin{align}
     \hat{\rho}^{W^n}_{\psi}=\hat{\rho}^{W^n}_{\hat{\pi}} \equiv \sum_{\mathbf{x}\in\mathcal{A}} p_{\mathbf{X}}\left(\mathbf{x}\right)\bigotimes_{i=1}^{n}\hat{\rho}^{W_i}_{\hat{\pi},x_i}\label{eq:willie-n-expansion}
\end{align}
where
$\hat{\rho}^{W_i}_{\hat{\pi},x_i} 
=\{\hat{\rho}^{W}_{\hat{\pi}} \text{ if }x_i=1; \hat{\rho}^{W}_{0} \text{ if } x_i=0\}$, with $\hat{\rho}^{W}_{\hat{\pi}}=\mathcal{N}_{A\to W}\left(\hat{\pi}\right)$ the non-innocent output state.

We upper-bound the QRE between Willie's output $\hat{\rho}_{\hat{\pi}}^{W^n}$ and the innocent state $\left(\hat{\rho}^W_0\right)^{\otimes n}$ as follows:
\begin{align}
D\left(\hat{\rho}^{W^n}_{\hat{\pi}}\middle\|\left(\hat{\rho}_0^W\right)^{\otimes n}\right)
&=   D\left( \left(\hat{\bar{\rho}}^{W}_{\hat{\pi}}\right)^{\otimes n}\middle\|\left(\hat{\rho}_0^W\right)^{\otimes n}\right)+o(1)  \label{eq:Mehrdad-bound}\\
    &= n D\left( \hat{\bar{\rho}}^{W}_{\hat{\pi}}\middle\|\hat{\rho}_0^W\right)+o(1)  \label{eq:QRE-additivity}\\
&\le q^2n D_{\chi^2}\left(\hat{\rho}^{W}_{\hat{\pi}}\middle\|\hat{\rho}_0^W\right), \label{eq:chi2-bound}
\end{align}
where $\hat{\bar{\rho}}^{W}_{\hat{\pi}}=(1-q)\hat{\rho}_0^{W} + q \hat{\rho}^{W}_{\hat{\pi}}$, \eqref{eq:Mehrdad-bound} is due to the adaptation provided in \iftoggle{isISIT}{\cite[Appendix A]{andersonAchievabilityCovertQuantum2025}}{Appendix~\ref{app:sparse-signaling-covertness}} of the covertness analysis for sparse signaling approach from 
\cite{tahmasbi21signalingcovert},
\eqref{eq:QRE-additivity} is from the additivity of the QRE over product states \cite[Ex.~11.8.7]{wilde16quantumit2ed}, and
\eqref{eq:chi2-bound} is by \cite[Lemma 1]{bullockCovertCommunicationClassicalQuantum2023} and \cite[Eq.~(9)]{ruskai1990convexity} for large enough $n$,
where the quantum $\chi^2$-divergence \cite{temme2010chi2} between two states $\hat{\rho}$ and $\hat{\sigma}$ is $D_{\chi^2}\left(\hat{\rho}\middle\|\hat{\sigma}\right) \equiv \tr\left(\hat{\rho}^2\hat{\sigma}^{-1}\right) - 1$. Thus, the right-hand side of the covertness requirement \eqref{eq:covert-req} is bounded by \eqref{eq:chi2-bound}. 
Assumptions in \eqref{eq:supports-covert-possible} and \eqref{eq:inf-chi-square-assumption} ensure $D_{\chi^2}\left(\hat{\rho}^{W}_{\hat{\pi}}\middle\|\hat{\rho}_0^W\right)$ is bounded for finite and infinite-dimensional output states, respectively.
Thus, Alice maintains covertness by choosing $q\leq c_\text{q}\sqrt{\frac{\delta_{\rm QRE}}{n}}$,
where \begin{align}
    c_\text{q}=\sqrt{\frac{1}{ D_{\chi^2}\left(\hat{\rho}^{W}_{\hat{\pi}}\middle\|\hat{\rho}_0^W\right)}} \label{eq:c_cov}.
\end{align}
\end{IEEEproof}

\emph{Remark:} Pauli twirling is often used to approximate noise present in non-Pauli channels. Indeed, we use this property to transform an arbitrary quantum channel into a Pauli channel for reliability analysis. However, Pauli twirling has an added benefit in covertness analysis, ensuring that, on average, the non-innocent input state appears as the maximally mixed state from Willie's point of view.
This not only simplifies the expressions but also leads to significant improvement in both $c_{\rm q}$ and $R$ for bosonic channels over the results in \cite{anderson2024covert-qce}.
This is because in \cite{anderson2024covert-qce} we added noise to ensure the Alice-to-Willie channel is entanglement-breaking, substantially reducing $R$.
Although Pauli twirling also has a deleterious effect on quantum communication, it is not as substantial.

\begin{figure*}[htb]
\centering
\includegraphics[width=\textwidth]{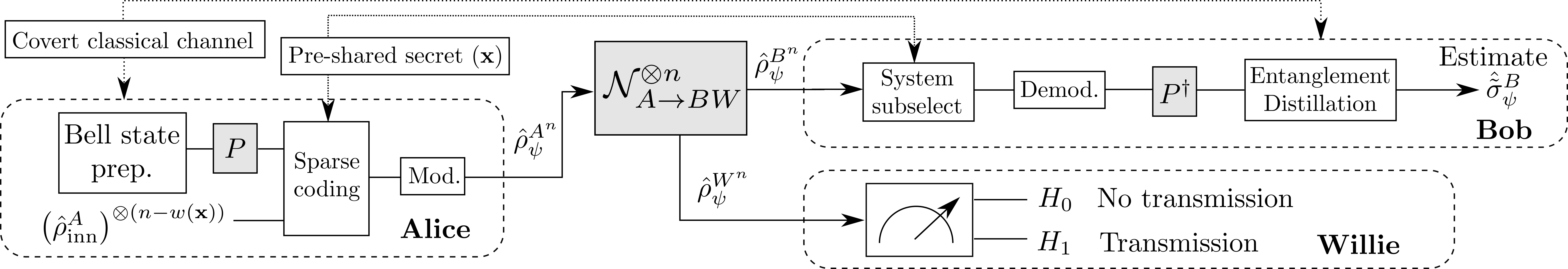}
\caption{Construction of achievability with the aid of a two-way covert classical channel. The construction is similar to Fig.~\ref{fig:construction-no-locc}, where instead of encoding a message using a QECC, Alice prepares $w(\mathbf{x})$ Bell states, performs sparse coding using the pre-shared secret $\mathbf{x}$, and transmits one-half of each to Bob. Bob still performs system subselection, and demodulation through projection. Alice and Bob use the classical covert communication link to perform entanglement distillation and teleportation to covertly transmit state $\hat{\sigma}^{A^\prime}_\psi$ to Bob. Notably, the sequence of Pauli gates, $\mathbf{y}$, no longer needs to remain secret.  }
\label{fig:construction-locc}
\end{figure*}

\subsection{Impact of Two-way Covert Classical Channel Assistance} 
\begin{theorem}
\label{thm:lower-bound-classical}
When a two-way covert classical channel is available, for $n$ large enough and arbitrary $\vartheta>0$, $M(n)\geq(1-\vartheta)\sqrt{n}c_{\mathrm{q}}R_{\mathrm{d}}\sqrt{\delta_{\text{QRE}}}$ qubits can be transmitted reliably and covertly over $n$ uses of an arbitrary quantum channel that acts independently on each channel use, where $q_\mathrm{c}$ is defined in \eqref{eq:c_cov}, $\delta_{\text{QRE}}$ is the QRE-covertness constraint, and $R_{\mathrm{d}}\geq (1-\pfail)\left[1-H(\vec{q}_{\text{tw}})\right]^+$ is the constant rate of entanglement distillation per channel use, with $\pfail$ and $\vec{q}_{\text{tw}}$ defined in the proof of Theorem \ref{thm:lower-bound}. The total number of covert classical bits required is at most $(1+\vartheta)\sqrt{n}c_{\mathrm{q}}\left(1+2R_{\mathrm{d}}\right)\sqrt{\delta_{\text{QRE}}}$.
\end{theorem}
\begin{IEEEproof}
Consider the construction in Fig.~\ref{fig:construction-locc}, which includes a two-way covert classical link. Instead of encoding a quantum message in a QECC, Alice prepares $w(\bf{x})$ Bell states and applies a random but public Pauli gate to one-half of the Bell state. She then applies the innocent state and sparse coding procedure described in the construction of the proof of Theorem \ref{thm:lower-bound}. Next, she modulates and transmits one-half of the Bell pair to Bob through quantum channel $\mathcal{N}_{A\to BW}$. Bob still performs system subselection, projection, and inverse of the random Pauli gate applied by Alice. However, instead of replacing the failed projection with the maximally mixed state, he transmits classically back to Alice a stream of $0$s and $1$s representing success or failure, respectively, for each system. This requires $w(\bf{x})\propto \sqrt{n}$ classical bits.

Alice and Bob perform hashing-based entanglement distillation on the $(1-\pfail)w(\bf{x})$ successful projections with an achievable rate for a Pauli error channel of $\left[1-H(\vec{q}_{\text{tw}})\right]^+$ \cite{bennet96qec}. Hence, with successful projection, they distill at a rate of $R_{\mathrm{d}} = (1-\pfail)\left[1-H(\vec{q}_{\text{tw}})\right]^+$. Alice then uses the covert classical link to transmit two classical bits per distilled Bell pair, performing teleportation of a desired state $\ket{\psi}^{A^\prime}$. By convexity, $R_d \ge R$. The total number of covert classical bits required is $w(\mathbf{x}) + 2w(\mathbf{x})R_{\mathrm{d}}$ where the first term represents Bob's communication of projection success or failure and the second is Alice transmitting two bits per distilled Bell pair. The upper bound on $w(\mathbf{x})\leq(1+\vartheta)\sqrt{n}c_{\mathrm{q}}\sqrt{\delta_{\text{QRE}}}$ yields the bound on the sufficient number of covert classical bits. Note that, although Alice and Bob must agree on a set of their noisy Bell pairs to measure in the hashing protocol, we do not transmit it on a covert classical channel as it can be pre-shared (it also does not have to be secret from Willie). 

We still require Alice and Bob to perform Pauli twirling, ensuring the channel with projection is Pauli. However, Alice preparing Bell states lifts the requirement of Willie being unaware of $\mathbf{y}$. This is because Willie already observes a maximally mixed state in each system when one-half of a Bell pair is transmitted through the channel. Thus, the covertness analysis is identical to \eqref{eq:willie-n-expansion} in the proof of Theorem \ref{thm:lower-bound}. 
\end{IEEEproof}

\section{Conclusion and Discussion} \label{sec:discussion}

We propose random-quantum-code-based schemes that achieve the SRL for covert quantum communication with and without a two-way classical covert channel.
We characterize the lower bounds for the constants in front of $\sqrt{n}$ for both schemes.
However, much is left for future investigation.
We expect challenges in deriving the converse for general quantum channels, as is typical in quantum information theory.
Thus, we will focus our study of the converse on specific yet widely-used channels, such as Pauli and erasure.
Notably, we have already derived a converse for covert quantum communication over a bosonic channel in \cite{anderson2024covert-qce}, although we believe that it can be improved.
Furthermore, we will attempt to reduce the size of the classical shared secret by adapting the resolvability techniques used in \cite{bullock2025fundamentallimitscovertcommunication}.
We will also investigate lifting the assumptions on Willie's output state for infinite-dimensional channels described in Section \ref{subsec:hypothesis-testing}.

Both quantum covert communication schemes in this work demodulate the received state by projecting it into the qubit space. The two-way classical covert communication allows distilling the entanglement from only the successful projections. This yields a significant advantage over the scheme without classical communication, where the effective channel combines Pauli and erasure channels. We replace unsuccessful projections (erasures) with maximally-mixed states. This enables the direct use of the hashing bound but does not employ the information about erasure locations. Indeed, stabilizer codes may correct up to twice as many erasure errors as Pauli errors \cite[Sec.~III.A]{grasslCodesQuantumErasure1997} with the erasure channel having a known capacity of $[1-2\pfail]^+$\cite{bennett1997capacities} where $\pfail$ is the probability of an erasure event. Inspired by the results for pure erasure channel \cite{bennett1997capacities,gottesman1997stabilizer}, we are investigating decoding strategies that use the erasure locations to correct both erasure and Pauli errors.

Finally, we would like to comment on the quantum resources required for implementing the proposed protocol, specifically the random stabilizer codes. Our results apply in the limit of asymptotically large codes. In both proposed protocols, this requires Alice and Bob to store large numbers of physical qubits in quantum memories and apply large Clifford circuits to them. Therefore, covert quantum communication needs to be investigated under practical constraints on quantum memory and circuit sizes.
We anticipate success using high-rate quantum codes, e.g., quantum low-density parity check (LDPC) codes~\cite{breuckmann2021quantum}, in addition to careful design of covert classical communication for the two-way scheme. 

\section*{Acknowledgement}
The authors are grateful to Mehrdad Tahmasbi for providing the details on the sparse coding analysis in \cite{tahmasbi21signalingcovert} \iftoggle{isISIT}{}{, which formed the basis for Appendix \ref{app:sparse-signaling-covertness}}. The authors also benefited from discussions with Matthieu R.~Bloch, Christos N.~Gagatsos, Brian J.~Smith, Ryan Camacho, Narayanan Rengaswamy, Kenneth Goodenough, and Saikat Guha. This material is based upon work supported by the National Science Foundation under Grants No. CCF-2006679 and EEC-1941583.

\iftoggle{isISIT}{
\newpage
}{}
\bibliographystyle{IEEEtran}
\bibliography{./papers.bib}

\appendices

\iftoggle{isISIT}{}{

\section{Covertness Analysis for Sparse Signaling} \label{app:sparse-signaling-covertness}

Here we adapt the approach from 
\cite{tahmasbi21signalingcovert} 
to show validity of \eqref{eq:Mehrdad-bound}. By the definition of QRE, we have:
\begin{IEEEeqnarray}{rCl}
\IEEEeqnarraymulticol{3}{l}{D\left(\hat{\rho}^{W^n}_{\hat{\pi}}\middle\|\left(\hat{\rho}_0^W\right)^{\otimes n}\right)}\IEEEnonumber\\
&=& D\left( \left(\hat{\bar{\rho}}^{W}_{\hat{\pi}}\right)^{\otimes n}\middle\|\left(\hat{\rho}_0^W\right)^{\otimes n}\right)\IEEEnonumber\\
&&+\left(S\left(\left(\hat{\bar{\rho}}^{W}_{\hat{\pi}}\right)^{\otimes n}\right)-S\left(\hat{\rho}^{W^n}_{\hat{\pi}}\right)\right)\IEEEnonumber\\
&&+\tr\left(\left(\left(\hat{\bar{\rho}}^{W}_{\hat{\pi}}\right)^{\otimes n}-\hat{\rho}^{W^n}_{\hat{\pi}}\right)\log\left(\left(\hat{\rho}_0^W\right)^{\otimes n}
\right)\right),\IEEEeqnarraynumspace\label{eq:qre_equality_decomposition}
\end{IEEEeqnarray}
where $S(\hat{\rho})=-\tr\left(\hat{\rho}\log\hat{\rho}\right)$ is the von Neumann entropy of quantum state $\hat{\rho}$.
We upper-bound the last two terms of \eqref{eq:qre_equality_decomposition}.
First, denote $\epsilon$ to be  
\begin{align}
    \epsilon&\equiv \frac{1}{2}\left\|\left(\hat{\bar{\rho}}^{W}_{\hat{\pi}}\right)^{\otimes n}-\hat{\rho}^{W^n}_{\hat{\pi}}\right\|_1,
\end{align}
and bound it as
\begin{align}
    \epsilon &=\frac{1}{2}\left\|\left(\hat{\bar{\rho}}^{W}_{\hat{\pi}}\right)^{\otimes n}-\hat{\rho}^{W^n}_{\hat{\pi}}\right\|_1
    \\\label{eq:data_processing}
    &\leq\frac{1}{2}\sum_{\mathbf{x}\in\{0,1\}^{n}}\left|\prod_{i=1}^{n}p_X(x_i)-p_{\mathbf{X}}(\mathbf{x})\right| 
    \\
    &=p\left(\bar{\mathcal{A}}\right) \\
    & \leq 2 \exp\left[-\frac{1}{3} qn\vartheta^2\right],\label{eq:chernoff}
\end{align}
where \eqref{eq:data_processing} is by the data processing inequality with classical statistical (total variation) distance, $p\left(\bar{\mathcal{A}}\right)=1-p\left(\mathcal{A}\right)$,
and the inequality in \eqref{eq:chernoff} is the Chernoff bound.
For our setting of $q\propto 1/\sqrt{n}$, $\epsilon$ decays to zero exponentially in $\sqrt{n}$.

Now, we must specialize our approach depending on the dimension of Willie's output states. 

\noindent{\bf Finite-dimensional output states:} Suppose that output states at Willie have dimension $d<\infty$. 
By Fannes--Audenaert inequality \cite[Th.~11.10.1]{wilde16quantumit2ed}, we have 
\begin{align}
    &\left(S\left(\left(\hat{\bar{\rho}}^{W}_{\hat{\pi}}\right)^{\otimes n}\right)-S\left(\hat{\rho}^{W^n}_{\hat{\pi}}\right)\right) \leq\epsilon n \log d + h(\epsilon),
\end{align}
where $h(x) = -x\log(x)-(1-x)\log(1-x)$ is the binary entropy function.
Note that \eqref{eq:chernoff} implies that this term vanishes as $n\to\infty$. 
Next, we bound the last term in \eqref{eq:qre_equality_decomposition}:
\begin{align}
    \tr&\left(\left(\left(\hat{\bar{\rho}}^{W}_{\hat{\pi}}\right)^{\otimes n}-\hat{\rho}^{W^n}_{\hat{\pi}}\right)\log\left(\left(\hat{\rho}_0^W\right)^{\otimes n}
\right)\right) \label{eq:finite-dim-holders} \\
&\phantom{=}\leq \left\|\left(\hat{\bar{\rho}}^{W}_{\hat{\pi}}\right)^{\otimes n}-\hat{\rho}^{W^n}_{\hat{\pi}}\right\|_1 \left\|\log\left(\left(\hat{\rho}_0^W\right)^{\otimes n}
\right)\right\|_\infty\\
&\phantom{=}=2\epsilon n \log\left(\frac{1}{\lambda_{\rm min}\left(\hat{\rho}^W_0\right)}\right),
\end{align}
where $\lambda_{\rm min}\left(\hat{\rho}^W_0\right)$ is the minimum eigenvalue of $\hat{\rho}^W_0$, and is non-zero by \eqref{eq:supports-covert-possible}. Noting that \eqref{eq:chernoff} implies that this term vanishes as $n\to\infty$ concludes the argument for finite-dimensional output states at Willie.

\noindent{\bf Infinite-dimensional output states:}  

Using the assumptions in Sec. \ref{subsec:hypothesis-testing}, we bound the second and third terms of \eqref{eq:qre_equality_decomposition} when Willie's output is infinite-dimensional. 

Let $\hat{H}^{W_i}$ be the Hamiltonian for the $i^{\text{th}}$ system at Willie.
Let $
E\equiv \max\left(\sum_{i=1}^n\tr\left(\hat{\bar{\rho}}^{W}_{\hat{\pi}}\hat{H}^{W_i}\right),\sum_{i=1}^n\tr\left(
\hat{\rho}^{W^n}_{\hat{\pi}}\hat{H}^{W_i}\right)\right)=O(n)$,
where $f(n)=O(g(n))$ means that $f(n)$ grows no faster asymptotically than $g(n)$: $\limsup_{n\to\infty} \left|\frac{f(n)}{g(n)}\right|<\infty$. By \cite[Lemma 15]{winter16cont},
\begin{align}
        &S\left(\left(\hat{\bar{\rho}}^{W}_{\hat{\pi}}\right)^{\otimes n}\right)-S\left(\hat{\rho}^{W^n}_{\hat{\pi}}\right)\leq2\epsilon n S_{\rm max}\left(\frac{E}{\epsilon n}\right) + h(\epsilon), \label{eq:infinite-dim-fannes}
\end{align}
where $S_{\rm max}(E)< \infty$ is the maximum entropy under energy constraint $E<\infty$. \cite[Rem.~13]{winter16cont}, \cite[Prop.~1(ii)]{Shirokov06entropychar} and the fact that $\epsilon n\to 0$ implies that \eqref{eq:infinite-dim-fannes} vanishes as $n\to\infty$. 

To bound the last term in \eqref{eq:qre_equality_decomposition}, we decompose $\left(\hat{\bar{\rho}}^{W}_{\hat{\pi}}\right)^{\otimes n}=p\left(\mathcal{A}\right)\hat{\rho}^{W^n}_{\hat{\pi}}+p\left(\bar{\mathcal{A}}\right)\label{eq:prob_not_in_A}\hat{\sigma}^{W^n}_{\hat{\pi}}$,
where $\hat{\sigma}^{W^n}_{\hat{\pi}}$ is a density operator with $E_{\hat{\sigma}}\equiv\sum_{i=1}^n\tr\left(\hat{\sigma}^{W^n}_{\hat{\pi}}\hat{H}^{W_i}\right)=O(n)$.
Let $\hat{\Delta}^{W^n}_{\hat{\pi}}\equiv \hat{\rho}^{W^n}_{\hat{\pi}}-\hat{\sigma}^{W^n}_{\hat{\pi}}$ and note that $\tr\left(\hat{\Delta}^{W^n}_{\hat{\pi}}\right)=0$ and $E_{\hat{\Delta}}\equiv\sum_{i=1}^n\tr\left(\hat{\Delta}^{W^n}_{\hat{\pi}}\hat{H}^{W_i}\right)=O(n)$.
Then,
\begin{IEEEeqnarray}{rCl}
\IEEEeqnarraymulticol{3}{l}
{\tr\left(\left(\left(\hat{\bar{\rho}}^{W}_{\hat{\pi}}\right)^{\otimes n}-\hat{\rho}^{W^n}_{\hat{\pi}}\right)\log\left(\left(\hat{\rho}_0^W\right)^{\otimes n}
\right)\right)}\IEEEnonumber\\
&=&p\left(\bar{\mathcal{A}}\right)\tr\left(\hat{\Delta}^{W^n}_{\hat{\pi}}\log\left(\left(\hat{\rho}_0^W\right)^{\otimes n}
\right)\right)\\
&=&p\left(\bar{\mathcal{A}}\right)\tr\left(\hat{\Delta}^{W^n}_{\hat{\pi}}\sum_{i=1}^n\log\left[\frac{1}{Z(\beta(E_0))}e^{-\beta(E_0) \hat{H}^{W_i}}\right]\right)\label{eq:operatorexp}\IEEEeqnarraynumspace\\
&=&p\left(\bar{\mathcal{A}}\right)n\log\left(\frac{1}{Z(\beta(E_0))}\right)\tr\left(\hat{\Delta}^{W^n}_{\hat{\pi}}\right) \notag \\
&\phantom{=}&-p\left(\bar{\mathcal{A}}\right)\beta(E_0)E_{\hat{\Delta}}\label{eq:qre_equality_third_term}\\
&=&o(1)\label{eq:qre_equality_fourth_term},
\end{IEEEeqnarray}
where $\eqref{eq:operatorexp}$ is due to operator exponentiation and the definition of $\hat{\rho}^W_0$ in \eqref{eq:rho0-Gibbs-form}, and \eqref{eq:qre_equality_fourth_term}
 follows from \eqref{eq:chernoff}.
}

\end{document}